# Crowd Simulation Modeling Applied to Emergency and Evacuation Simulations using Multi-Agent Systems

João E. Almeida[1,2], Rosaldo Rosseti[1,2], António Leça Coelho[3]

[1]LIACC – Laboratório de Inteligência Artificial e Ciência de Computadores
[2]FEUP – Faculdade de Engenharia da Universidade do Porto
[3]LNEC – Laboratório Nacional de Engenharia Civil
Rua Dr. Roberto Frias s/n, 4200-465 Porto, Portugal
joao.almeida@engenheiros.pt

**Abstract.** In recent years crowd modeling has become increasingly important both in the computer games industry and in emergency simulation. This paper discusses some aspects of what has been accomplished in this field, from social sciences to the computer implementation of modeling and simulation. Problem overview is described including some of the most common techniques used. Multi-Agent Systems is stated as the preferred approach for emergency evacuation simulations. A framework is proposed based on the work of Fangqin and Aizhu with extensions to include some BDI aspects. Future work includes expansion of the model's features and implementation of a prototype for validation of the propose methodology.

## 1 Introduction

Crowd and group simulations are becoming increasingly important in the computer games industry and in emergency simulation. Applications range from the entertainment to more serious use like pedestrian behavior in the real world or in panic situations. This paper summarizes a synthesis of what has been done in recent years in this field, discussing the various aspects involved, from social sciences to the computer implementation of modeling and simulation using Multi-Agent Systems. A framework is proposed based on the work of Fangqin and Aizhu with extensions to include some BDI aspects. Future work includes expansion of the model's features and implementation of a prototype for validation of the propose methodology.

### 1.1 Crowd behavior

Studying crowd behavior in emergency situations is difficult since it often requires exposing real people to the actual, possibly dangerous, environment. Fire drills (fig.1) are a possible approach but hardly recreating the truly panic conditions, people tend to take it not seriously. A good computational tool that takes into consideration the human and social behavior of a crowd could serve as a viable alternative.

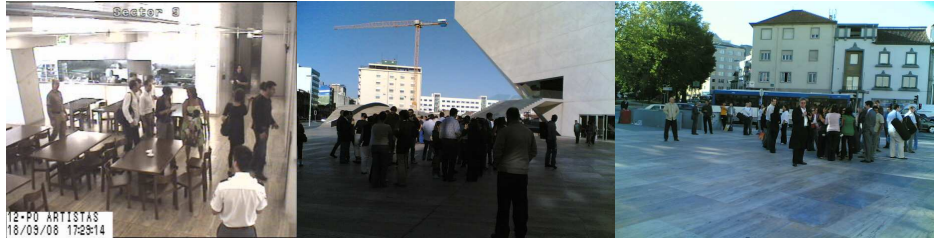

**Fig. 1.** During a fire drill at "Casa da Musica"[1] Oporto.

Computer models for emergency and evacuation situations have been developed and most research into panics has been of empirical nature and carried out by researchers from social sciences [1],[3],[5],[6].

### 1.2 Normal pedestrian behavior

Pedestrian crowds have been empirically studied for the past decades [1],[2]. The evaluation methods applied were based on direct observation, photographs, and time-lapse films. Apart from behavioral investigations, the main goal of these studies was to develop computer animated realistic applications, for the game industry, design elements of pedestrian facilities, or planning guidelines for architectural building and urban design.

In their common environment pedestrians tend to show some basic attributes. For example people always try to find the shortest and easiest way to reach their destination. If possible they avoid detours, even if the shortest way is crowded. The basic principle is the "least effort principle", which means everyone tries to reach their goal as fast as possible spending the least amount of energy and time.

### 1.3 Individual and crowd behavior in emergency and panic situations

Most of the normal behavior vanishes when pedestrians face an emergency situation (it does not always have to be an emergency situation, similar effects can be observed for example in crowds trying to get the best seats at a concert or consumers running for sales). Observations made for pedestrian crowds in emergency situations feature typically the same patterns. As people try to leave the building as fast as possible, the desired velocity increases which leads to some characteristic formations. As nervousness increases there is less concern about comfort zone and finding the most convenient and shortest way.

---

[1] Photos taken by the author in 2008

It is observable, for example, that if people have to leave a building in an emergency situation and they don't know the structure of the building well enough, they would run for the exit they used as an entrance, even if other exits might be easier to reach or even safer.[1]

They also might lose the ability to orient themselves in their surrounding and thus show herding or flocking behavior [3]. Not only do they lose certain abilities, they also start to exhibit new behaviors like pushing or other physical interactions. Nonadaptive crowd behaviors are recognized to be responsible for the death and injury of most victims in crowd disasters. Nonadaptive crowd behaviors refer to the destructive actions that a crowd may experience in emergency situations, such as stampede, pushing, knocking, and trampling on others.

### 1.4 Herding or Flocking

Herding tries to describe a human group dynamics visible in emergency situations (fig.2). When people get nervous and feel panic, they lose the ability to act logically and to decide on their own. As a result of this lack of independence, people tend to follow others in the assumption they could get them out of the dangerous area. On one side this could actually help people to escape faster, but if for example smoke is reducing the visibility or the person leading the group does not know the structure of the building well enough, it could also reduce the chance to find an exit. So instead of people wandering around on their own, more and more flocks of people start to form with increasing anxiety or nervousness. As simulations have shown none of the extremes (people walking around on their own or as a single large group) results in optimal evacuation time [1],[5].

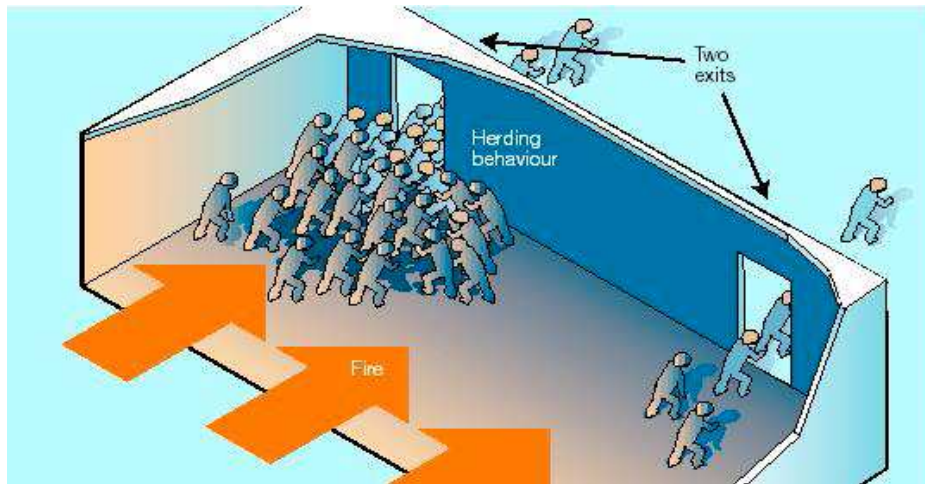

**Fig. 2.** Crowd trying to escape from smoke-filled room [1].

## 1.5 Arching and Clogging

Observations have shown a phenomenon called arching, which appears when a big crowd with a high desired velocity tries to pass through a door. Instead of passing through the door in less time, or giving the oncoming pedestrians a chance to pass through the door, the door gets clogged and the crowd gets arch-shaped (fig.3).

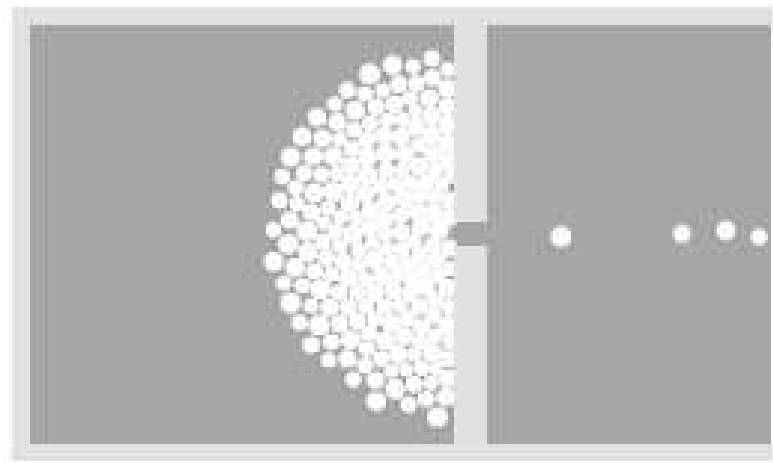

**Fig. 3.** Arching and clogging [1].

## 2 Related work

There are three main reasons for developing computer simulation for crowd behaviors: first to test scientific theories and hypotheses; second, to test design strategies; third, to create phenomena about which to theorize [8]. A full understanding of crowd behaviors would require exposing real people to the specific environment for obtaining empirical data, which is difficult since such environments are often dangerous in nature. In addition to studying crowd behavior based on observations and historical records, computer simulation is a useful alternative that can provide valuable information to evaluate a design, to help the planning process, and for dealing with emergencies.

Human behaviors are complex emergent phenomena, which are difficult to capture into computers as mathematical equations. There are several techniques to model crowds. Existing models can be categorized into one of the following groups.

## 2.1 Flow-based modeling

Flow-based models use the density of nodes in continuous flows. The basic principle is the analogy with fluid and particle motions. Often are called *macroscopic models*. Characteristics are defined beforehand thus all particles behave in the same way.

In this kind of models the simulated physical environment is defined as a network of nodes. The nodes represent physical structures, such as rooms, stairs, lobbies, and hallways that are all connected and comprise a single structure from which an evacuation is executed. The nodes contain people. Certain nodes are designated as destination nodes identifying the possible exits. For each node, the usable area must be calculated and allowance is made for the presence of closets, equipment, and other such items, as well as the space which persons place between themselves and a wall. Besides nodes, the model also requires the provision of specification for arcs. Arcs are passageways between building components with two variables: traversal time or the amount of time it takes to cross the passageway, and an arc flow capacity which delimits the amount of human occupants that can cross the passageway per unit time. One example of this type of modelling is EVACNET4 [4],[18].

## 2.2 Cellular Automata

In this kind of modeling space is discretized. A matrix is created plotting areas in a two-dimensional array. The simulation technique uses a time-frame pre-defined in which the occupants can move from one position or node to another, assuming it is free or it is not an obstacle. Each element can have several values: empty, occupied by a person, occupied by some object, or part of the limits (wall). The movement occurs at every step of the defined time-frame when occupants can move to one of the adjacent nodes. Each person can only move to an empty node, and directions are limited to the eight possible nearby nodes. Microscopic and macroscopic analyses are both permitted.

This type of model is simple to implement but fails when trying to replicate the erratic movement of people in real life, since only limited movement is allowed. Also it is not easy to model different speeds and interaction between people, due to the grid shape of space. Nevertheless this is the most used type for crowd modeling in games and more serious applications. One popular example is Exodus in early versions [4]. Another example is EGRESS [8].

## 2.3 Agent-based

Multi-Agent Systems (MAS) approach to this problem is probably the most realistic solution since it allows to model each individual person with their own unique characteristics, but related with all surrounding persons, thus recreating the real world interactions among human beings.

SIMULEX [4] was the first application to use MAS. Exodus latest's versions and PedGO also use MAS [9].

## 3 Multi-Agent Systems in Evacuation Simulators

In recent years MAS has been used as the preferred method to simulate crowd movement in different scenarios [7],[8],[9],[11],[16]. The enormous complexity of agent modeling, the need of data and rules to feed the system and the computational time needed (although according to Moore's Law computers' processing power keeps increasing) have created some difficulties to this approach.

However, investigation is going on and new papers describing work in this field are becoming more and more common. The possibilities offered by MAS are immense, as long as social rules and interaction knowledge among people is known and fed to the model. Social knowledge from researchers of other fields besides modeling and computational areas are welcome.

### 3.1 MAS Model

The model must be as complete as possible with all variables supplied to the virtual environment and then made available to the agents.

Human individuals are modeled as autonomous agents who interact with a virtual environment and other agents according to the individual's characteristics (which may vary from person to person) using global rules derived from the world where the system is created. Each agent has a limited vision of the world. Depending on the environment and the behavioral levels of individuals and their relationships with the group (or the crowd), the agent could interact and react in a collaborative or competitive manner. In contrast to agent-based systems for design applications, there is no global system control in the simulation model. In fact, the objective here is to be able to observe the random dynamics among the individuals (agents) in the simulation environment. To simulate human cognitive processes agents continuously sense and assess the surrounding environment making decisions based on their own decision model. The crowd social behaviors are collectively observed as emergent phenomena.

### 3.2 BDI Agents

MAS can use different levels of complexity and implement social-like behavior, using the BDI technique (*Beliefs, Desires, Intentions*) where agents are driven by *Desires* (the goals), according to certain *Beliefs* (set of knowledge of the world) and *Intentions* (actions) to fulfill the *Desires*. For instance, in an emergency evacuation simulation, agents' *Desires* are to leave the place where they are, due to fire or other hazard, as quickly as possible, using the fastest and safest path (following the *Beliefs*) and taking the necessary actions (*Intentions*).

Social forces such as comfort zone, pushing and fighting for space, should also be modeled and interactions between agents tested and validated.

BDI agents will implement more complex decision making processes and interaction among them can help scientists find new relations and derive modeling mathematical rules to understand crowd behavior in normal and emergency situations. This would help designers to build safer buildings, planners to prepare better emergency plans and educators to find the best strategies for emergency plans. Although much work has been done, much more is needed to achieve realistic results.

## 4 A Framework for Crowd Simulation in Emergency Situations

In order to implement a good framework for crowd simulation in emergency situations, the steps to follow are:
- select the best methodologies to implement;
- use the best modeling data available;
- use of MAS for crowd modeling;
- create an open-source framework ready to accept new add-ons;
- use modular development, OO languages and off-the-shelf technology;
- allow easy inclusion of newer algorithms.

### 4.1 Model structure

The model structure proposed by Fangqin and Aizhu (fig.4) is a good starting point. This model proposes the use of readymade and available software thus saving much developing time.

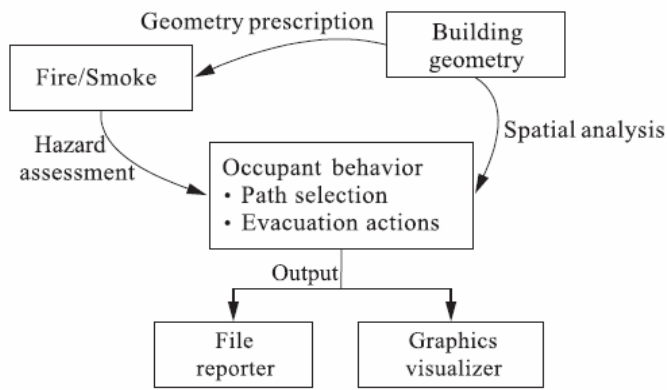

**Fig. 4.** Model flowchart [7].

*Building geometry* and *Fire/Smoke* models are based on existing and tested software, like FDS [19],[20] and PyroSim, a commercial software package, for 3D CAD building design [21]. Data interchange can be done using file systems and batch processing, since computational time needed is high and real-time visualization can only happen after all calculations are complete.

Fire Dynamics Simulator (FDS) is a Computer Fluid Dynamics (CFD) software developed by NIST[2] for fire hazard assessment, freely available for scientific use[19],[20]. FDS uses geometry data based on a database structure shared by PyroSim [21].

The *Occupant behavior* module implements crowd modeling based on geometry from PyroSim and hazard data from FDS. Its main objective is to define the exit path for each occupant, considering all interactions between the agents and environment.

### 4.2 Agent's attributes

For the agent modeling, the attributes are described bellow (as shown on fig. 5):

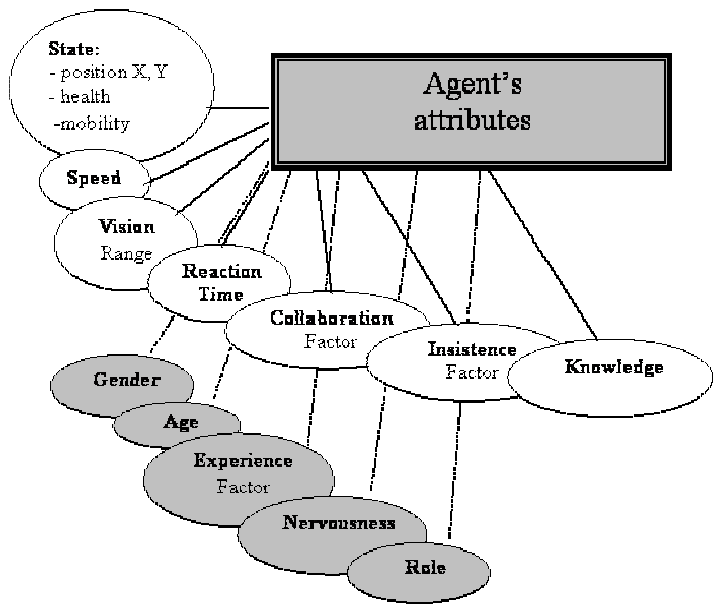

**Fig. 5.** Agent's attributes as proposed by Fangqin and Aizhu [7] and extensions (shaded items).

**State.** Physical data of the agent: current position (X,Y), being X and Y the coordinates of the discrete spatial location in the virtual space; health condition [0-1] real interval showing if the agent is alive (1), injured or ill ( 0 < health condition < 1) or dead (0); mobility (0-can't move; 1-normal behavior; 2-panic behavior).
**Speed.** Agent speed velocity (m/s) varies with health condition and mobility, ranging from 0 (when health=0 or mobility=0) to 7 m/s (when running in panic).

---
[2] NIST: National Institute of Standards and Technology, USA

**Vision.** Vision is the visible range from current location. Depends on health condition and OD[3] given by FDS. Also must be able to detect obstacles and other agents.
**Reaction Time.** Reaction Time (RT) is the time an occupant spends to decide what action should take. Typically this time ranges from few seconds to some minutes [6]. This attribute will be in seconds.
**Collaboration.** Some degree of cooperation among occupants is typically refers in all studies concerning this area [5],[6],[12],[13],[14]. The attributes here will reflect the path selection algorithm that will be implemented.
**Insistence.** The insistence factor defined on the interval [0,1] indicates the probability of maintaining the current evacuation strategy. When an agent is experiencing low evacuation efficiency, the attribute decreases and leads to strategy adjustments.This attribute will be used to adjust the path selection algorithm.
**Knowledge.** Represents the degree of familiarity of the occupant with the building. This factor varies with the knowledge of the surroundings and will increase when the agent gets more acquainted with the space.

Other attributes to implement a BDI architecture are proposed in this paper to expand the model possibilities:

**Gender.** Many studies refer differences between men and women reactions in panic or stress situations [5],[6],[12],[13].
**Age.** Another important factor determining behavior according referred studies.
**Experience.** Previous experience in exercises or real situations are proved to be important in the decision making process [14]. In this attribute factors such as (1) knowledge to use fire extinguishers (2) participation in fire drills (3) previous experience in emergency situations, should be taken into account.
**Nervousness.** Factor indicating the degree of nervousness or anxiety, of the agent when facing emergency situations.
**Role.** The agent role can be set as a coefficient related to the importance in evacuation scenarios. Can be a hierarchical status (director/chief/simple employee or teacher/student) indicating the importance of his/her decisions and influence for the surrounding agents. This attribute can be used to implement the leader-follower model some researchers propose with dynamic grouping [10]. Integer (0: none; 1: top level; 2: $2^{nd}$ level; … n: $n^{th}$ level).

### 4.3 Occupant behavior module

This model implements the interactions between agents and the environment. Hazard information is received from the FDS module with all variables related to temperature, smoke, pressure, toxicity of air and visibility, in each of the compartments or spaces in the scenario. Geometry is supplied from the same database used by FDS and designed using PyroSim.

For the *occupants behavior* MAS will be used. The complexity of the decision making process will depend upon agent's attributes and environment data supplied the *Building geometry* and *Fire/Smoke* models. Rules can vary from simple reaction action to more complex BDI with interaction between agents.

---

[3] Optical Density: unit 1/m, measures the visibility in smoke condition

The actions agent's will take into account the following aspects:
1) analysis of environment conditions (alarm, temperature, presence of smoke, etc.) and determine the need of evacuation (this will give Reaction Time);
2) observation of other agent's behavior and eventually follow their actions (depends on agent's level of hierarchy);
3) visibility conditions, knowledge of the environment, surrounding exits;
4) knowledge of the building and nearest path to safe place or exit;
5) presence of obstacles or other people clogging exits;
6) physical conditions;
7) social forces when crowd is forcing to pass through a clogged exit.

Within this context, the decision making algorithm should provide: (1) an exit path (2) adapt route whenever conditions change (3) inform other agent's of actions taken.

To improve the model, works related with social forces and human interaction should be used, like the recent studies of Moussaïd et al [17] where the decision process for pedestrians and behavior rules are mathematically modeled based on empirical observations.

### 4.4 Output module

The occupant behavior module actions will be the input for this module. These actions can be either saved in file, for later processing, or directly sent to graphic display using 3D software like OpenGL.

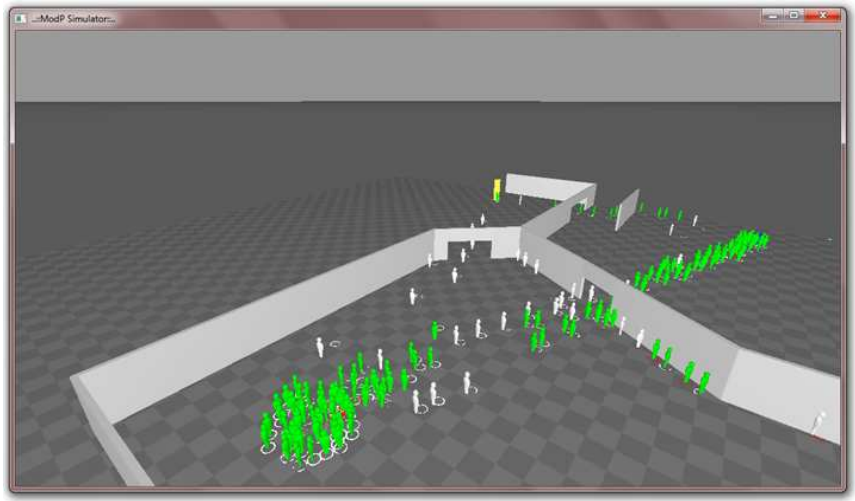

**Fig. 6.** Modp 3D viewer [22],[23].

Features to implement related to the animation, could be: selecting the camera position; detailing regarding the human representation; background textures; visualization of smoke and flames, etc. Exodus shows a good approach to realistic images.

One possibility would be using ModP [22],[23] for 3D pedestrian visualization module (fig.6).

## 5 Conclusions and future work

This paper introduces the need of crowd simulation for computer games or more serious applications like emergency evacuation. Problem overview is described including some of the most common techniques used. Multi-Agent Systems approach is stated as the preferred technique for emergency evacuation simulations. A framework for crowd simulation in emergency situations is proposed based on the work of Fangqin and Aizhu [7] with some extensions to include some BDI agent's aspects such as (1) sensors for the real world, (2) social forces and (3) interaction with other agents.

Future work includes expansion of the model's features and implementation of a prototype for validation of the propose methodology.